\documentclass[%
 reprint,
 amsmath,amssymb,
 aps,
pra,
]{revtex4-1}

\usepackage{graphicx}
\usepackage{dcolumn}
\usepackage{bm}
\usepackage{epstopdf}
\usepackage{siunitx}

\begin{document}

\preprint{APS/123-QED}

\title{Temperature dependence of the Kerr nonlinearity and two-photon absorption in a silicon waveguide at \SI[detect-weight=true]{1.55}{\micro \meter}}

\author{Gary F. Sinclair}
\email{gary.f.sinclair@bristol.ac.uk}
\author{Nicola A. Tyler}
\author{D\"{o}nd\"{u} Sahin}
\author{Jorge Barreto}
\author{Mark G. Thompson}

\affiliation{%
Quantum Engineering Technology Labs, School of Physics, H.H. Wills Physics Laboratory, University of Bristol, Tyndall Avenue, Bristol BS8 1TL, United Kingdom}%

\date{\today}

\begin{abstract}
We measure the temperature dependence of the two-photon absorption and optical Kerr nonlinearity of a silicon waveguide over a range of temperatures from \num{5.5} to \SI{300}{\kelvin} at a wavelength of \SI{1.55}{\micro \meter}.  The two-photon absorption coefficient is calculated from the power dependent transmission of a \SI{4.9}{\pico \second} pulse.  We observed a nearly two-fold decrease in the two-photon absorption coefficient from \SI{0.76}{\centi \meter/\giga \watt} at \SI{300}{\kelvin} to \SI{0.42}{\centi \meter/ \giga \watt} at \SI{5.5}{\kelvin}.  The Kerr nonlinearity is inferred from the self-phase modulation induced spectral broadening of the transmitted pulse.  A smaller reduction in Kerr nonlinearity from \SI{5.2d-18}{\meter^2 / \watt} at \SI{300}{\kelvin} to \SI{3.9d-18}{\meter^2 /  \watt} at \SI{5.5}{\kelvin} is found.  The increased ratio of Kerr to absorptive nonlinearity at low temperatures indicates an improved operation of devices that make use of a nonlinear phase shift, such as optical switches or parametric photon-pair sources.  We examine how the heralding efficiency of a photon-pair source will change at low temperature.  In addition, the modelling and experimental techniques developed can readily be extended to other wavelengths or materials of interest.
\end{abstract}

\maketitle

\section{Introduction}
The high material nonlinearity and tight mode confinement present in silicon waveguides offers a useful platform for constructing integrated nonlinear photonic devices.  A variety of nonlinear photonic applications have been demonstrated in silicon, including optical frequency combs \cite{Griffith2015}, all-optical switching \cite{Lacava:13}, all-optical logic \cite{Xu:07}, parametric amplification \cite{Foster2006} and parametric photon-pair sources \cite{Sharping:06, Faruque:18} for quantum information processing.  Driven by these applications, there has been a considerable effort to accurately measure  \cite{doi:10.1063/1.2737359, doi:10.1063/1.2750523, 0268-1242-23-6-064007} and model \cite{1242372, 4628477, doi:10.1063/1.3570654} the nonlinear optical properties of silicon.  Although these studies have focused on room temperature measurements, some investigations of the temperature dependence of the one- and two-photon indirect absorption coefficient at a wavelength of \SI{1.06}{\micro \meter} \cite{PhysRevLett.30.901, 0022-3719-12-18-029} and free-carrier lifetime \cite{Pernice:11} have been performed.  At this wavelength, a strong temperature-dependence of the one-photon indirect absorption coefficient has been observed, showing two orders of magnitude change over the \SI{100}{\kelvin} to \SI{300}{\kelvin} range \cite{0022-3719-12-18-029}.  However, no study of the temperature dependence of the Kerr nonlinearity and two-photon absorption has been performed at wavelengths around \SI{1.55}{\micro \meter} in the telecom C-band, as commonly used in most integrated silicon-photonic devices.

One application area that is likely to require cryogenic operation is integrated quantum photonics.  Waveguide-integrated single-photon detectors are essential for measurement-based quantum computing.  To achieve high-efficiency, low-noise detection along with fast control feedback, the monolithic integration of superconducting nanowire single photon-detectors with photonic circuits is required \cite{doi:10.1063/1.3657518, Pernice2012}.  Operation of the chip at few-Kelvin temperatures will potentially impact on the performance of nonlinear photonic components, such as parametric photon-pair sources, where the brightness, purity and heralding efficiency depend on the nonlinear optical parameters of the waveguide \cite{PhysRevA.94.063855, Husko2013}.  In particular, the heralding efficiency of a photon-pair source is known to be adversely affected by the nonlinear cross two-photon absorption present in silicon at wavelengths in the the telecom C-band \cite{Husko2013}.  Additionally, silicon photonics also offers a promising platform for chemical and biological sensing in the mid-infrared wavelength range where the fundamental vibrational modes of most chemical bonds are present \cite{Soref2010, Zou:18}.  However, at longer wavelengths background thermal noise becomes an ever larger constraint on device performance \cite{Soref2010, Barh:18}, suggesting that low-temperature operation could be of benefit.

In this article, we determine the Kerr nonlinearity and two-photon absorption (TPA) present in a silicon waveguide between the temperatures of \num{5.5} and \SI{300}{\kelvin} at a wavelength of \SI{1.551}{\micro \meter}.  By modeling the propagation of a \SI{4.9}{\pico \second} optical pulse along a \SI{19.09}{\milli \meter} waveguide, we can relate the observed nonlinear transmission and pulse spectral broadening to the nonlinear waveguide parameters.  This is the first measurement of both the TPA and Kerr nonlinearity at cryogenic temperatures in the technologically important telecom C-band.  In addition, although the techniques developed have been used to characterize a silicon waveguide at \SI{1.551}{\micro \meter}, they could more generally be applied to other third-order nonlinear materials and at other wavelengths.

\section{Modeling pulse propagation}
\begin{figure}[t]
\includegraphics{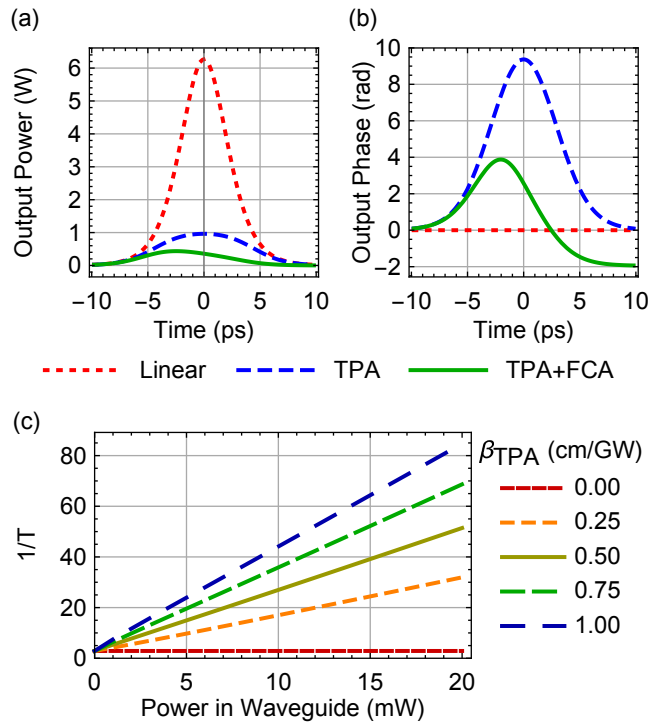}
\caption{\label{fig:model} (a) The modeled output power and (b) phase of a \SI{4.9}{\pico \second} pulse with peak in-waveguide power of \SI{18.0}{\watt} (\SI{5}{\milli \watt} average) and room-temperature nonlinear parameter values.  (c) The modeled inverse transmission of a pulse for a range of two-photon absorption parameters and fixed room-temperature FCA, as a function of the time-averaged input power at the start of the waveguide.}
\end{figure}

We begin by constructing a model of pulse propagation through a nonlinear optical waveguide.  The propagation of an optical pulse is commonly described by the {\em Nonlinear Schr\"{o}dinger Equation}~\cite{Yin:07},
\begin{equation}
\label{eqn:nlse}
\frac{\partial E}{\partial z} = \left (i k_o n_2 - \frac{\beta_\mathrm{TPA}}{2} \right ) \vert E \vert^2 E - \frac{\sigma_\mathrm{FCA}}{2} (1 + i \mu) N_c E - \frac{\alpha}{2} E,
\end{equation}
where $E$  is the electric field envelope (in units of \si{\sqrt{\watt / \meter^2}}), $k_o=2 \pi/\lambda$ is the wavenumber, $n_2$ is the Kerr coefficient, $\beta_\mathrm{TPA}$ is the TPA coefficient, $\sigma_\mathrm{FCA}$ is the free-carrier absorption (FCA) coefficient, $\mu$ determines the relative strength of the free-carrier dispersion (FCD), $N_c$ is the free-carrier density and $\alpha$ is the linear loss.  The generation of free carriers at each point of the waveguide is governed by,
\begin{equation}
\frac{\partial N_c}{\partial t} = \frac{\beta_\mathrm{TPA}}{2 \hbar \omega} \vert E \vert^4 - \frac{N_c}{t_c},
\label{eqn:dndt}
\end{equation}
where $t_c$ is the free-carrier lifetime.  An exact closed-form solution of these propagation equations can be found when free-carrier effects can be neglected, as occurs when the pulse energy and repetition rate are sufficiently low~\cite{Yin:07}.  Alternatively an approximate solution can also be found when free-carrier effects only weakly modify the pulse shape \cite{4840394}.  In our work we numerically solve the propagation equations, as detailed in Appendix~\ref{app:mpp}.  Briefly, the propagation equations are solved using a first-order finite-difference method~\cite{agrawal2007nonlinear}, assuming excitation of the quasi transverse-electric (quasi-TE) mode of a $220 \times 500$ \si{\nano \meter} silicon-on-insulator waveguide of length \SI{19.09}{\milli \meter}.  Figure \ref{fig:model} shows (a) the output pulse power and (b) the phase, as predicted by our model for room temperature values of the nonlinear parameters and a high-power input pulse (\SI{5}{\milli \watt} in-waveguide power).  At this power, the short duration (\SI{4.9}{\pico \second}) and low repetition rate of our optical pulses ensures that TPA makes the largest contribution to the nonlinear loss.  However, the effect of free-carriers is equally as important as the Kerr nonlinearity for the phase, where the self-phase modulation (SPM) and FCD act with opposite signs.  This causes a reduction in the peak phase shift and an asymmetry in the phase profile, due to the accumulation of free-carriers over the duration of the pulse.  Free-carrier effects are included in the model, using a temperature-dependent model for the FCA parameter (see Appendix~\ref{app:mpp}).

Following earlier works \cite{PhysRevLett.30.901, 0022-3719-12-18-029, 0022-3727-12-3-012} we monitor the inverse transmission through the waveguide ($1/T = P_\mathrm{in} / P_\mathrm{out}$), where the time-averaged power is $P=\Gamma \int_\mathrm{pulse} P(\tau) d \tau$, $\Gamma$ is the laser repetition rate (\SI{50}{\mega \hertz}) and the time integral is taken over the input or output pulse as appropriate.  Figure \ref{fig:model}(c) shows the modeled inverse transmission along the waveguide (not including coupling losses) for a range of TPA coefficients ($\beta_\mathrm{TPA}$) and a fixed room-temperature value for FCA (see Appendix \ref{app:mpp}).  When $\beta_\mathrm{TPA}=0$ there is no nonlinear absorption and the inverse transmission is a constant, equal to the inverse of the linear propagation loss along the waveguide (\SI{2.4}{\deci \bel / \centi \meter}).  The inverse transmission is seen to remain almost a straight line with an increasing slope for larger values of the TPA parameter.  Plotting the data in this way has the benefit of decoupling the two fit parameters: the intercept, from which the linear loss is deduced and the slope, from which the TPA parameter is found.  This leads to more accurate fits when analyzing noisy data.

\section{Experimental Setup}
\begin{figure*}[t]
\includegraphics{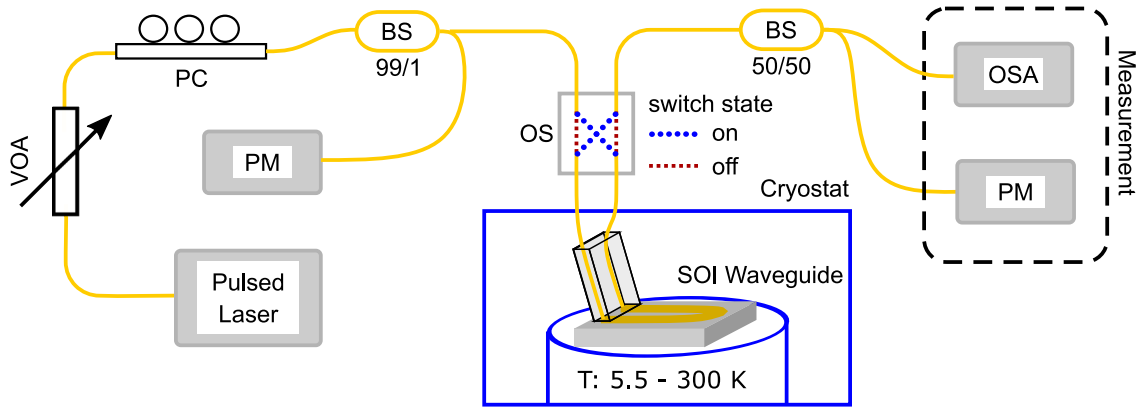}
\caption{\label{fig:setup}The experimental setup used when measuring the transmission and spectral broadening of pulses transmitted through a silicon waveguide: VOA, variable optical attenuator; SMF, single-mode fiber; PC, polarization controller; BS, fiber beam-splitter; OS, 2 $\times$ 2 optical cross-over switch; OSA, optical spectrum analyser; PM, power meter; SOI, silicon-on-insulator.}
\end{figure*}
Figure \ref{fig:setup} depicts the setup used to take measurements of the transmission and spectrum of an optical pulse after propagation through a \SI{19.09}{\milli \meter} waveguide at various temperatures between \num{5.5} and \SI{300}{\kelvin}.  Optical pulses of \SI{4.9}{\pico \second} FWHM (full-width half-maximum) were generated by a PriTel Femtosecond Fiber Laser, with a repetition rate of \SI{50}{\mega \hertz}, \SI{196}{\watt} peak power and wavelength of \SI{1551.8}{\nano \meter}.  The pulse duration was measured using an intensity autocorrelator (Femtochrome Research FR-103PD), assuming a $sech^2$ profile, as expected for a passively mode-locked fiber laser \cite{Thomson:98}.  A variable optical attenuator (Oz Optics, DA-100) was used to step the power linearly between \SI{30}{\deci \bel} and \SI{0}{\deci \bel} of attenuation.  At each attenuation setting (20 in total) the optical spectrum and transmission were recorded (Anritsu MS9740A and Thorlabs S154C powermeter respectively).  Before each power scan the transmission was maximised using a polarization controller (Agilent 11896A) to ensure a consistent polarization state (quasi-TE) in the waveguide.  Power was monitored at the input using a 99:1 fiber coupler and at the output using a 50:50 splitter to divide the light equally between the output power meter and optical spectrum analyser.  The splitting ratio and insertion loss of each coupler were measured to ensure correct estimates of the input/output power could be made.  Before coupling light on/off the chip a $2 \times 2$ cross-over switch was used to allow light to be propagated in either direction along the waveguide.  Any asymmetry in the efficiency of the on-chip grating couplers can be compensated for by taking measurements in both orientations, as discussed in Appendix \ref{app:bi}.  This is possible because the degree of nonlinear absorption by the waveguide depends on if the input grating coupler constitutes a larger or smaller proportion of the overall loss.  This provides a method of unambiguously determining each grating coupler efficiency independently.  At each temperature setting power scans were taken in pairs: initially with the light propagating along the waveguide in one direction and then in the reverse direction.  Multiple pairs of measurements were taken for each temperature setting to monitor the reproducibility of our data.

The chip under test was cooled in a Lakeshore Cryogenic Probe Test Station (Model CPX), which was modified to allow vertical coupling to and from the chip via focused grating couplers ($\approx$ \SI{5}{\deci \bel} loss per coupler).  This enabled cooling of the chip to a minimum of \SI{5.5}{\kelvin}, maintaining a temperature stability of at least \SI{0.1}{\kelvin}.  Between each temperature setting the system was typically allowed 15 minutes to reach thermal equilibrium and steady coupling.
\section{Results}
\subsection{Two-photon absorption}
\begin{figure}[b]
\includegraphics{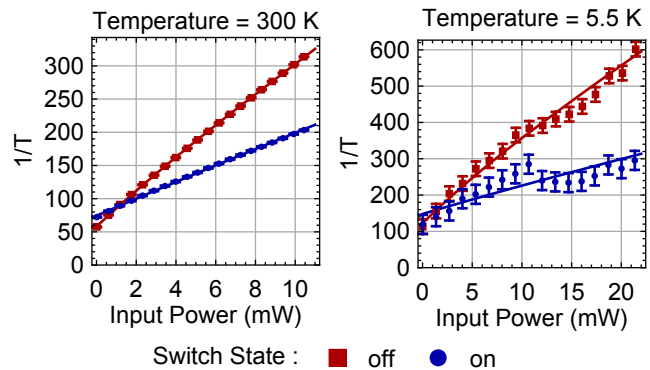}
\caption{\label{fig:tpadata}Model fits to the measured inverse transmission (fiber to fiber) at temperatures of \SI{300}{\kelvin} and \SI{5.5}{\kelvin}. Data are taken with light propagating in both directions along the waveguide, as indicated by the cross-over switch being in the ``on" or ``off" state.}
\end{figure}
Using the model developed above, the inverse transmission data for each power scan was fitted to extract the TPA coefficient.  Figure \ref{fig:tpadata} shows example fits at \SI{5.5}{\kelvin} and \SI{300}{\kelvin}.  Due to asymmetry of the grating couplers a larger apparent nonlinearity is observed in the ``off" direction, as indicated by the larger gradient of the fit.  This occurs because, given the same overall fiber-to-fiber linear loss, if the input grating coupler has a lower loss than the output then the larger power in the waveguide will give rise to a larger nonlinear response.  By taking data in both orientations, the correct TPA coefficient can be found from the geometric mean of the pair (see Appendix \ref{app:bi}).  We note that the cross-over switch introduces a slight excess loss when in the ``on" state, as indicated by the larger value of $1/T$ at the zero-power intercept, although this does not affect the fitted TPA value.  Data at low temperatures suffers from greater noise due to fluctuations in the coupling caused by a mechanical resonance of the cryostat sample stage.  This was particularly noticable at temperatures below \SI{50}{\kelvin}.  To compensate for this a larger number of power scans were taken at lower temperatures than at higher (at least 20 scans at each temperature below \SI{50}{\kelvin}, and at least 6 scans at each temperature above \SI{50}{\kelvin}).  In total 432 pairs of power scans were taken across the entire temperature range.

\begin{figure}[t!]
\includegraphics{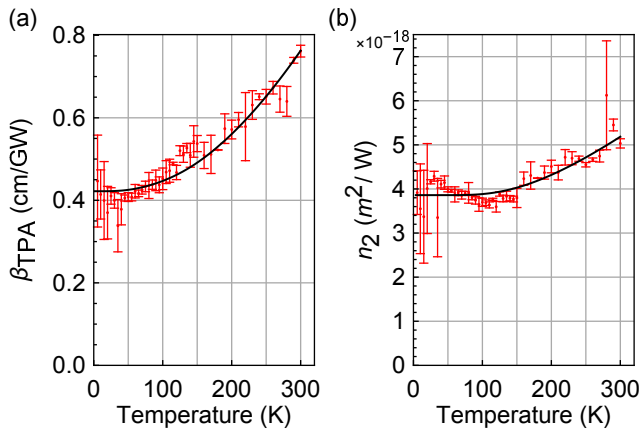}
\caption{\label{fig:tpan2} (a) The two-photon absorption as a function of temperature, fitted using Eq. (\ref{eqn:TPA}).  Based on the fitting, TPA is seen to decrease by $45\%$ from 300 to \SI{5.5}{\kelvin}. (b) The Kerr nonlinearity, fitted using Eq. (\ref{eqn:n2}).  A smaller reduction in the Kerr nonlinearity of 25\% is observed over the same temperature range.}
\end{figure}
Figure \ref{fig:tpan2}(a) shows the fitted TPA coefficient extracted from the inverse transmission power scans.  Each point in the graph is the arithmetic mean of the TPA coefficient determined from several pairs of transmission measurements taken in both directions, where the error bars are the standard deviation of the individual pair-wise measurements.  We fit our data by following \cite{4628477} and assuming that the dominant contribution to the TPA comes from phonon assisted two-photon allowed-allowed transitions across the lowest indirect bandgap.  In addition, we include contributions from both transverse acoustic ($E^\mathrm{TA}_\mathrm{ph}/k_B =  \SI{212}{\kelvin}$) and transverse optical ($E^\mathrm{TO}_\mathrm{ph}/k_B =  \SI{670}{\kelvin}$)  phonon  branches \cite{doi:10.1063/1.356496}:
\begin{equation}
\label{eqn:TPA}
\beta_\mathrm{TPA} = E_\mathrm{gap}(T)^{3/2} \sum_{b} \left \{ K_b \left [F^b_+(T) + F^b_-(T) \right ] \right \}.
\end{equation}
Here, $b \in \{\mathrm{TA}, \mathrm{TO} \}$ sums over the phonon branches and $F^b_\pm$ arise due to terms involving the creation (+) or annihilation (-) of a phonon in the corresponding branch:
\begin{eqnarray}
F^b_+(T) &=& \frac{(2 \hbar \omega - E_\mathrm{gap}(T) - E^b_\mathrm{ph})^2}{\exp (E^b_\mathrm{ph}/k_B T) - 1 },\\
F^b_-(T) &=&  \frac{(2 \hbar \omega - E_\mathrm{gap}(T) + E^b_\mathrm{ph})^2}{1-\exp (-E^b_\mathrm{ph}/k_B T)},
\end{eqnarray}
where, $\hbar \omega$ is the photon energy (\SI{0.797}{\eV}) and $E_\mathrm{gap}(T) = E_\mathrm{gap}(0) - \beta T^2/(T + \delta)$ is the temperature dependent bandgap, where the values $E_\mathrm{gap}(0) = \SI{1.156}{\eV}$, $\beta = \SI{7.021d-4}{\eV \per \kelvin}$, and $\delta = \SI{1108}{\kelvin}$ are taken from \cite{VARSHNI1967149}.  As is apparent from Fig. \ref{fig:tpan2}(a) the temperature dependence of the TPA is well represented by (\ref{eqn:TPA}), where we have found the fit parameters $K_\mathrm{TA}=0.233$ and $K_\mathrm{TO}=2.138$.  We note that the largest contribution to the temperature dependence ($\sim$ 80 \%) comes from the phonon population, rather than the change in the bandgap.  Since the two-photon transition is phonon assisted, the transition probability is seen to decrease at lower temperatures as the phonon population declines.  However, transitions are always possible even in the absence of a significant thermal bath of phonons, as a phonon can be created to assist the transition.  This is in contrast to one-photon indirect transitions at \SI{1.06}{\micro \meter}, where the rapidly changing density of state close to the band edge leads to a much larger temperature dependence of the absorption \cite{0022-3719-12-18-029}.

\subsection{Kerr nonlinearity}
\begin{figure}[h]
\includegraphics{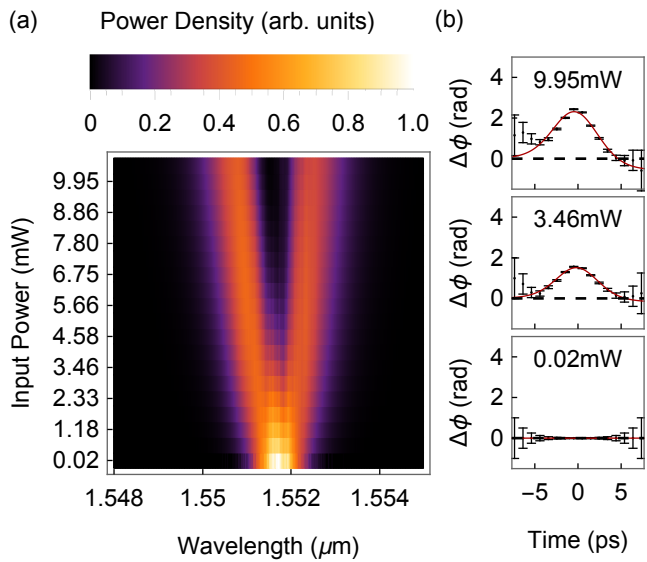}
\caption{\label{fig:spectrum}(a) The measured power spectrum at the output of the waveguide.  Each power spectrum (horizontal cross-section of the plot) is normalized to the same total pulse energy for clarity.  (b) The reconstructed temporal phase ($\Delta \phi$) of the output pulse for three different input powers: (black points) - data, (red line) - fitted numerical model}
\end{figure}

To determine the Kerr nonlinearity, the output spectrum of the transmitted pulse was measured at a range of input powers, as shown in Fig. \ref{fig:spectrum}(a).  As expected, SPM causes a broadening of the pulse spectrum as the input power is increased.  A very slight blue-shift of the central wavelength can also be seen due to FCD.  For each input power the temporal phase of the output pulse was reconstructed using the Saxton-Gerchberg phase retrieval algorithm, full details of which can be found in \cite{Fienup:82}.  This algorithm requires knowledge of the output power spectrum, which is straightforward to measure using an optical spectrum analyzer, and the pulse envelope in the time-domain.  To estimate the pulse shape at the output of the waveguide, we used the transmission data (discussed above) to determine the degree to which the pulse was distorted due to the nonlinear absorption from the initial pulse profile.  This estimated output pulse shape was used in the algorithm to reconstruct the pulse phase.  We note however, that the reconstructed phases are quite insensitive to the exact shape of the temporal pulse envelope, as the observed spectral broadening is largely due to the nonlinear phase imparted to the pulse, rather than a slight distortion of the pulse envelope \cite{Yin:07}.

A selection of reconstructed phase profiles and fits to our numerical pulse evolution model are shown in Fig. \ref{fig:spectrum}(b).  All reconstructed phases are baseline corrected against the lowest power (\SI{0.02}{\milli \watt}) scan, which is why this appears as a flat phase profile.  Although the input pulse was not in fact transform limited, the additive nature of the accumulated phase means that the exact input phase profile is not important, so long as the temporal pulse length is correctly known.  In addition, due to the significant length of optical fiber in our setup, it was necessary to subtract the background phase accumulated due to SPM in the optical fibers.  Removing this background contribution to the nonlinear phase is necessary, otherwise this would lead to an over-estimate of the Kerr nonlinearity in the waveguide.  To characterize the background Kerr nonlinearity, a series of transmission spectra through the chip were measured as the coupling to the chip was gradually decreased.  The spectral broadening was seen to decrease as the power in the waveguide dropped.  Once the fiber-to-fiber excess loss had been increased to approximately \SI{20}{\deci \bel} the output spectrum was seen to remain constant as the chip was further decoupled.  This indicated that pulse propagation in the chip was of sufficiently low power so as to be in  the linear regime, whereas the nonlinearity produced in the fibers leading up to the chip could still be measured.  A power scan was taken at this decoupling to characterize the Kerr-nonlinearity of the input fibers, allowing us to subtract this background contribution to the phase from all of our subsequent measurements when the chip was well-coupled.  No significant further nonlinearity was expected in the output fibers, due to the lower power present after the chip.  The reconstructed phases shown in  Fig. \ref{fig:spectrum}(b) have been background corrected using the method described.  As can be seen, the reconstructed phases clearly exhibit a profile consistent with SPM due to the Kerr nonlinearity and a slight asymmetry due to FCD.  We use the numerical model developed to fit the phase profiles and extract the Kerr nonlinearity for each of the reconstructed phase profiles.

Figure \ref{fig:tpan2}(b) shows the measured Kerr coefficient as a function of the sample temperature.  Although some temperature dependence is observed, the decrease in Kerr nonlinearity at low temperatures ($25\%$) is seen to be a little over half the ($45\%$) decline observed in the absorptive nonlinearity.  We choose to fit the data to an empirical model motivated by that used for the TPA coefficient:
\begin{eqnarray}
\label{eqn:n2}
n_2(T) &=& n_2(0) \big [ \frac{1}{\exp (E_\mathrm{ph}/k_B T) - 1}  \nonumber \\
&& +  \frac{1}{ 1 - \exp(-E_\mathrm{ph}/k_B T)} \big ].
\end{eqnarray}
Here, the fit parameters are the Kerr coefficient at \SI{0}{\kelvin}, $n_2(0)=\SI{3.86d-18}{\meter^2 \watt^{-1}}$, and a parameter $E_\mathrm{ph}/k_b=\SI{576}{\kelvin}$, which has the form of a phonon energy and is somewhat close to the value of the transverse optical phonon energy.  We stress that this model is purely empirical and has been chosen to match the data in the absence of any other known models for the predicted temperature dependence of the Kerr nonlinearity.

\section{Nonlinear Figure of Merit}
In many photonic applications, TPA acts as a parasitic nonlinear loss mechanism, reducing the amount of useful Kerr nonlinearity that can be achieved.  For instance, it has been shown that TPA places a limit on the losses of an all-optical switch \cite{Mizrahi:89}, and that this loss can be further compounded at high switching rates by subsequent free-carrier effects \cite{Lacava:13}.  Similarly, it has been proposed that cross two-photon absorption (XTPA) between pump and signal/idler photons in a parametric photon-pair source acts as a fundamental limit to the heralding efficiency of such sources \cite{Husko2013}.  For this reason, it can be useful to introduce the dimensionless figure of merit $\mathrm{FOM} = n_2/(\lambda \beta_\mathrm{TPA})$ to describe the ratio of the useful Kerr nonlinearity to the parasitic TPA \cite{Mizrahi:89, lin2007dispersion}.  Here, we briefly investigate how the heralding efficiency of a parametric photon-pair source can be understood to depend on the nonlinear FOM.

As an illustration, we consider the simplest possible photon-pair source: a single-mode, non-dispersive, silicon waveguide.  For this type of source, photon pairs are generated over a broad bandwidth and narrow-band filtering is applied to the herald photon to improve the purity of the heralded photons.  It has been shown that the probability of generating one photon-pair per pulse can be written as \cite{PhysRevA.94.063855},
\begin{equation}
p_\mathrm{pair} = \frac{(\gamma L P)^2}{2} \sqrt{\frac{1 - \mathcal{P}^2}{\mathcal{P}}},
\label{eqn:ppair}
\end{equation}
where $\gamma = (k_0 n_2)/A_\mathrm{eff}$ is the waveguide nonlinear parameter, $A_\mathrm{eff}$ is the effective area of the waveguide, $P$ is the pump peak power, $L$ is the waveguide length and $\mathcal{P}$ is the purity of the heralded photons.  Here, the source is assumed to be operating in the weakly pumped regime, so that higher-order terms can be neglected.  If the only loss mechanism present is XTPA then the heralding efficiency (also known as the Klyshko efficiency \cite{0049-1748-7-5-A15, Rarity:87}) will be limited only by the probability of XTPA between the pump and heralded photons \cite{Husko2013}.  Therefore, the heralding efficiency will be equal to,
\begin{equation}
\eta_\mathrm{heralding} = \frac{1}{(1 + \xi)^2},
\end{equation}
where $\xi = \alpha_\mathrm{TPA} L P$ is a dimensionless parameter describing the nonlinear loss and $\alpha_\mathrm{TPA} = \beta_\mathrm{TPA}/A_\mathrm{eff}$ is the waveguide nonlinear absorption parameter.  Using (\ref{eqn:ppair}) and the definition of the nonlinear FOM we find we can express $\xi$ in terms of the source purity, brightness and material FOM:
\begin{equation}
\xi = \frac{1}{2 \pi \mathrm{FOM}} \sqrt{2 p_\mathrm{pair} \frac{\mathcal{P}}{\sqrt{1- \mathcal{P}}}}.
\end{equation}
Evaluating the heralding efficiency for some reasonable parameters ($p_\mathrm{pair} = 0.05, \mathcal{P}=0.9$) we find a room temperature heralding efficiency of 0.74, improving to 0.79 at \SI{0}{\kelvin} using FOMs of 0.44 and 0.59 respectively (based on values given in Table \ref{tab:values} of Appendix \ref{app:mpp}).  Thus, cryogenic cooling to \SI{0}{\kelvin} is seen to have a moderately positive effect on heralding efficiency.  We note that the nonlinear FOM has been seen to improve significantly at longer wavelengths towards the mid-infrared.  This has led to suggestions to develop integrated silicon photonic components at longer wavelengths.  For instance, a room-temperature nonlinear FOM of 4.4 has been measured at \SI{2.2}{\micro \meter} in \cite{lin2007dispersion}.  Using this value for the FOM, a heralding efficiency of 0.97 is achieved, demonstrating the importance of the FOM for determining parametric source performance.  Nonetheless, we note that this simple source design has not been optimized for heralding efficiency, and other designs may exhibit bettern performance without the need to operate at longer wavelengths.

\section{Conclusions}
We have measured the temperature dependence of the TPA and Kerr nonlinearity at a wavelength of \SI{1.55}{\micro \meter} in the technologically important telecom C-band.  TPA was found to decrease by $45\%$ from 300 to \SI{5.5}{\kelvin}, whereas a smaller reduction of $25\%$ was observed for the Kerr nonlinearity.  The near two-fold reduction in TPA contrasts with the near vanishing of the one-photon indirect absorption previously observed at \SI{1.06}{\micro \meter} \cite{0022-3719-12-18-029, PhysRev.111.1245}.  The significantly stronger temperature dependence at the shorter wavelength appears to be due to the very rapidly changing density of states close to the band edge, whereas at \SI{1.55}{\micro \meter} the reduction in absorption is primarily due to the reduction in phonon population.

The stronger reduction in TPA as compared to Kerr nonlinearity, leads to an improved nonlinear figure of merit (FOM) at low temperatures.  We discuss how the nonlinear FOM is an important metric that determines the performance of many nonlinear photonic devices, such as all-optical switches and parametric photon-pair sources.  In particular, we examine how the heralding efficiency of a simple waveguide parametric photon-pair source depends on the nonlinear FOM, and show a moderate improvement in heralding efficiency at low temperatures.

When characterizing the nonlinearity of our silicon waveguide, several techniques have been employed that may be applicable to other nonlinear integrated photonic experiments.  In particular, the use of bi-directional measurements to eliminate coupling-loss uncertainty and phase-retrieval of the measured spectra to visualize the time-domain nonlinear effects, may be of general use to the community.

\begin{acknowledgments}
This work was supported by the Engineering and Physical Sciences Research Council (EPSRC) under the grant EP/L024020/1.  M.G.T. acknowledges fellowship support from EPSRC grant EP/K033085/1.  The authors would also like to thank Gerardo E. Villarreal-Garcia for assistance with cryogenic equipment.
\end{acknowledgments}

\appendix
\section{\label{app:mpp}Modeling pulse propagation}
A comprehensive account of modeling nonlinear optical phenomena in silicon waveguides is provided in \cite{Lin:07}.  However, some of the more immediately relevant fundamentals from that paper will be outlined below for convenience.  Equation (\ref{eqn:nlse}) describes the propagation of an optical pulse, where the electric field $E=E(\vec{r},\tau)$, $\vec{r}$ is the position vector, $\tau=t-z/v_g$ is the retarded time (where propagation is along the z-axis and $v_g$ is the group velocity), and the electric field is measured in units of \si{\sqrt{\watt \per \meter^2}}.  It is convenient to factorize the electric field such that $E = F(x, y) A(z, \tau)$, where $F(x,y)$ is the dimensionless transverse mode profile ($ F \in [0, 1]$) and $A(z, \tau)$ describes evolution along the propagation axis.  However, in most waveguide experiments it is the optical power, rather than the intensity, that is measured.  We therefore choose to introduce $\bar{A}(z, \tau) = (\int \vert F(x,y) \vert^2 dx dy)^{1/2} A(z,\tau) $, where $\bar{A}$  has units of \si{\sqrt{\watt}} and $\vert \bar{A} \vert^2$ corresponds to the power integrated over the entire mode cross-section.  Doing so, we find the propagation equation, 
\begin{equation}
\label{eqn:abar}
\frac{\partial \bar{A}}{\partial z} = \left (i \gamma -\frac{\alpha_\mathrm{TPA}}{2} \right ) \vert \bar{A} \vert^2 \bar{A} - \frac{\sigma_\mathrm{FCA}}{2} (1 + i \mu) \bar{N}_c \bar{A} - \frac{\alpha}{2} \bar{A},
\end{equation}
where $\gamma = (k_0 n_2)/A_\mathrm{eff}$ and $\alpha_\mathrm{TPA} = \beta_\mathrm{TPA}/A_\mathrm{eff}$ are the nonlinear parameters for the waveguide, $A_\mathrm{eff}=\left [ \int \vert F(x,y) \vert^2 dx dy \right ]^2/ \int_\mathrm{core} \vert F(x,y) \vert^4 dx dy$ is the effective mode area for SPM (where the lower integral is taken over the nonlinear core only) and other quantities are as defined in Eq. (\ref{eqn:nlse}).  The free-carrier density again has units of \si{\meter^{-3}}, but is averaged over the mode cross-section:
\begin{equation}
\bar{N}_c(z, t) = \frac{\int N_c(x,y,z,\tau) \vert F(x, y) \vert^2 dx dy}{\int \vert F(x, y) \vert^2 dx dy}.
\end{equation}
By doing so, free carriers are appropriately weighted according to their spatial overlap with the transverse mode in the waveguide.  The free-carrier generation equation becomes:
\begin{equation}
\label{eqn:nbar}
\frac{\partial \bar{N}_c}{\partial t} = \frac{\bar{\beta}_\mathrm{TPA}}{2 \hbar \omega} \vert \bar{A} \vert^4 - \frac{\bar{N}}{\tau_c},
\end{equation}
and $\bar{\beta}_\mathrm{TPA} = \beta_\mathrm{TPA} \int_\mathrm{core} \vert F(x,y) \vert^6 dx dy / \left [ \int \vert F(x,y) \vert^2 dx dy \right ]^3$ is weighted to take account of the mode distribution during the generation process and the subsequent interaction with the field (see Eq. (36) in \cite{Lin:07}).  The transverse mode profile (fundamental quasi-TE mode) was found using a commercial mode solver (Lumerical Mode Solutions).  The system of coupled differential equations (\ref{eqn:abar}) and (\ref{eqn:nbar}) could now be solved as one-dimensional propagation equations, where all details of the transverse modal structure have been absorbed into the definition of the parameters $\gamma$ and $\bar{\beta}_\mathrm{TPA}$.  However, a further simplification can be made by noting that rather than calculating the complex field amplitude, the phase can be calculated from the solution for the power and free-carrier distribution.  That is, if we take $\bar{A}(z, \tau) = \sqrt{P(z, \tau)} \exp [i \phi(z, \tau)]$ then from Eq. (\ref{eqn:abar}) we find,
\begin{equation}
\label{eqn:dpdz}
\frac{\partial P}{\partial z}=- \alpha_\mathrm{TPA} P^2 - \sigma_\mathrm{FCA} \bar{N}_c P - \alpha P,
\end{equation}
and the phase of the pulse at the output of the waveguide can be calculated from:
\begin{equation}
\phi(L,\tau) =  \gamma \int^L_0 P(z, \tau) dz - \frac{\sigma_\mathrm{FCA} \mu}{2} \int^L_0 \bar{N}_c(z, \tau) dz.
\end{equation}
Therefore, it is only necessary to numerically calculate the pulse power and free-carrier distribution as a function of position and time, for the range of $\gamma$ and $\sigma_\mathrm{FCA}$ values expected and input powers used.  This is in contrast to solving Eq. (\ref{eqn:abar}) directly, which would have require numerical solutions for the full range of Kerr and FCD parameters also.  Thus when fitting the reconstructed phases (see Fig. \ref{fig:spectrum}.(b)), the much simpler semi-analytic phase Ansatz can be used:
\begin{equation}
\phi(L,\tau) = \gamma S(\tau) - \frac{\sigma_\mathrm{FCA} \mu}{2} T(\tau),
\end{equation}
where $S(\tau) =  \int^L_0 P(z, \tau) dz$ and $T(\tau) =  \int^L_0 \bar{N}_c(z, \tau) dz$ need to be numerically calculated for the full range of TPA and FCA parameters expected and input powers used.  We note that when using the phase Ansatz to fit the reconstructed phase profiles, that it is only $\gamma$ and $\mu$ that are used as fit parameters: the in-waveguide power and TPA parameter had already been deduced from the transmission measurements for each temperature setting.

A temperature-dependent model of FCA was used to model the expected transmission and output phase-profiles.  A suitable model for FCA is given by \cite{1051012,Sun2013}
\begin{equation}
\label{eqn:fca}
\sigma_\mathrm{FCA} = \frac{q_e^3 \lambda^2}{4 \pi^2 \epsilon_0 c^3 n} \left ( \frac{1}{m_e^{*2} \mu_e} +\frac{1}{m_h^{*2} \mu_h} \right ),
\end{equation}
where $q_e$ is the electron charge, $\epsilon_0$ the permitivity of free space, $c$ the speed of light, $n=3.48$ the refractive index of silicon, $m^*_e=0.3 m_0$ and $m^*_h=0.4 m_0$ are the electron and hole effective masses expressed in terms of the free-electron mass, $m_0$, and $\mu_e$ and $\mu_h$ are the electron and hole mobilities.  Although the carrier effective masses are in general temperature dependent, they are not expected to change significantly over the range of temperatures in our study \cite{Riffe:02}.  Rather, the dominant contribution to the temperature dependence arises through the electron and hole mobilities.  The empirical models for carrier mobility developed in \cite{1482195} (see Eqn. (8) and (13) therein) show a temperature- and density-dependence of the carrier mobility.  In our study, carrier-densities are typically in the region of $10^{18}$ \si{\centi \meter^{-3}}, which results in a nearly constant mobility, and therefore constant $\sigma_\mathrm{FCA}$, from \SI{300}{\kelvin} down to \SI{50}{\kelvin}.  Below \SI{50}{\kelvin}, the carrier mobility is seen to increase rapidly, resulting in a rapid decrease in FCA as the temperature approaches \SI{0}{\kelvin}.

Table \ref{tab:values}  provide a selection of values for the nonlinear parameters used in our analysis as a function of temperature.  The values for $\alpha_\mathrm{TPA}$ and $n_2$ are those produced by the fitting functions (\ref{eqn:TPA}) and (\ref{eqn:n2}) respectively, whereas $\sigma_\mathrm{FCA}$ values are from the FCA model (\ref{eqn:fca}) and were used as an input to our numerical model.
\begin{table}[h]
\begin{center}
\begin{tabular}{*{3}{c@{\hspace{6mm}}}c}
\hline \hline
T  & $\alpha_\mathrm{TPA}$ & $n_2$  &  $\sigma_\mathrm{FCA}$\\
(\si{\kelvin}) &  (\si{\centi \meter \per \giga \watt}) & (\si{\nano \meter^2 \watt^{-1}}) &  ($10^{-22}$ \si{\meter^2}) \\
\hline
300 & \num{0.761} & \num{5.18} & \num{3.7}\\
150 & \num{0.492} & \num{4.03} &  \num{3.1}\\
50 & \num{0.424} & \num{3.86} &  \num{2.7} \\
5.5 & \num{0.420} & \num{3.86}  & \num{0.9} \\
0.0 & \num{0.420} & \num{3.86}  & \num{0.0} \\
\hline \hline
\end{tabular}
\end{center}
\caption{Temperature dependence of the nonlinear parameters.}
\label{tab:values}
\end{table}

\section{\label{app:bi}Correcting for coupling asymmetry}
When measuring the fiber-to-fiber transmission though an on-chip waveguide in the linear-optical regime, it is impossible to independently determine the coupling losses.  However, uncertainty in our knowledge of the on-chip power constitutes an important source of error when performing nonlinear measurements, as the measured nonlinear response varies quadratically with our estimate of the on-chip power.  This can lead to an incorrect estimate of the nonlinear parameters.  However, by taking pairs of transmission measurements with light propagating in both directions it is possible to determine the correct coefficients.  We show this by considering only the TPA below, although this can be generalized in a straightforward way to include free-carrier effects as well.  Consider the propagation equation (\ref{eqn:dpdz}), where the only nonlinearity considered is TPA:
\begin{equation}
\frac{\partial P}{\partial z} =  - \alpha_\mathrm{TPA} P^2 - \alpha P.
\end{equation}
We can introduce a rescaled power $p(z)$ such that $P(z, \tau) = P(0, \tau) p(z)$, and $p(0)=1$ at the start of the waveguide.  Then,
\begin{equation}
\frac{\partial p}{\partial z} = -A p^2 - \alpha P,
\end{equation}
where $A = \alpha_\mathrm{TPA} P(0, \tau)^2$ and we write the solution as $p=p(z; A)$.  By fitting this solution to our transmission data, the coefficient $A$ can be found, although we note that individually $\alpha_\mathrm{TPA}$ and $P(0, \tau)$ are still unknown.  However, taking a pair of measurements in both orientations will result in a pair of A-coefficients, $A' = \alpha_\mathrm{TPA} P'(0)$ and $A'' = \alpha_\mathrm{TPA} P''(0)$, where the different powers at the start of the waveguide arise due to the asymmetric coupling losses.  The waveguide has two grating couplers for coupling to external fibers, the losses of which we arbitrarily denote as $\eta_L$ and $\eta_R$.  We can therefore express the product of the A-coefficients as:
\begin{equation}
A' A'' = (\alpha_\mathrm{TPA} \eta_L P_\mathrm{in}) ( \alpha_\mathrm{TPA} \eta_R \sqrt{\eta_X} P_\mathrm{in}),
\end{equation}
where $P_\mathrm{in}$ is the power just before the input grating coupler and $\eta_X$ is the total excess-loss that is introduced by our cross-over switch when in the ``on" state.  However, when fitting each individual transmission power scan, we have no way of knowing $\eta_L$ and $\eta_R$ individually.  Therefore, it is necessary to assume symmetric losses on both grating couplers when performing the fitting.  For this reason we have,
\begin{equation}
A' A'' = (\alpha'_\mathrm{TPA} \sqrt{\eta_L \eta_R} P_\mathrm{in}) ( \alpha''_\mathrm{TPA} \sqrt{\eta_L \eta_R \eta_X} P_\mathrm{in}),
\end{equation}
where $\alpha'_\mathrm{TPA}$ and $\alpha''_\mathrm{TPA}$ are the apparent TPA coefficients that we extract from each individual fitting and $\sqrt{\eta_L \eta_R}$ is the average grating coupler loss that we are forced to use during the fitting.  Comparing these expressions it is straightforward to show that the real TPA parameter can be calculated from the geometric mean of the apparent TPA parameters found from each individual fit:
\begin{equation}
\alpha_\mathrm{TPA} = \sqrt{\alpha'_\mathrm{TPA} \alpha''_\mathrm{TPA}}
\end{equation}
This can be generalized in a straightforward manner to include free-carrier effects, showing $\sigma_\mathrm{FCA} = \sqrt{\sigma'_\mathrm{FCA} \sigma''_\mathrm{FCA}}$.  Similarly, by evaluating $A'/A''$ we find $\alpha'_\mathrm{TPA}/ \alpha''_\mathrm{TPA} = \eta_L / \eta_R$, from which the individual coupling losses can be deduced:
\begin{equation}
\eta^2_L = \sqrt{\eta_L \eta_R} \left(\frac{\alpha'_\mathrm{TPA}}{\alpha''_\mathrm{TPA}} \right ), \quad \eta^2_R = \sqrt{\eta_L \eta_R} \left ( \frac{\alpha''_\mathrm{TPA}}{\alpha'_\mathrm{TPA}} \right ).
\end{equation}

\bibliographystyle{apsrev4-1}
\bibliography{lowtemp}

\end{document}